\documentclass{mem}
\usepackage{natbib}
\usepackage{txfonts}
\usepackage{balance}
\usepackage{graphicx}
\usepackage[a4paper]{hyperref}
\idline{79}{1}
\begin{document}
\def\teff{$T\rm_{eff }$}
\def\kms{$\mathrm {km s}^{-1}$}

\title{
The $M_\bullet - \sigma_\ast$ Project
}

   \subtitle{}

\author{
D. Batcheldor\inst{1} 
          }

  \offprints{D. Batcheldor}

\institute{
Center for Imaging Science
Rochester Institute of Technology
Rochester, NY, 14623, USA. 
\email{dan@astro.rit.edu}
}

\authorrunning{Batcheldor}

\titlerunning{The $M_\bullet - \sigma_\ast$ Project}

\abstract{
There is an intimate link between supermassive black hole (SMBH) mass ($M_\bullet$) and the stellar velocity dispersion ($\sigma_\ast$) of the host bulge. This has a fundamental 
impact on our understanding of galaxy and SMBH formation and evolution. However, the scatter, slope and zero-point of the relation is a subject of some debate. For any progress 
to be made on this relation, the established values of $M_\bullet$ and $\sigma_\ast$ must be robust. Over 50\% of current $M_\bullet$ estimates have been made using the technique 
of stellar dynamics. However, there is serious concern over this method that prompts their re-evaluation. In addition, it is not clear how best to define $\sigma_\ast$. The aim 
of the $M_\bullet - \sigma_\ast$ Project is to use STIS long-slit spectroscopy, integral field spectroscopy and the latest stellar models, to best estimate the values of $M_\bullet$ 
and $\sigma_\ast$ in as many cases as possible. The project will determine the most appropriate properties of the $M_\bullet - \sigma_\ast$ relation itself. 
}
\maketitle{}

\section{Introduction}

The discovery of the relationship between supermassive black hole (SMBH) mass ($M_\bullet$) and the stellar velocity dispersion ($\sigma_\ast$) of the host bulge 
\citep{2000ApJ...539L...9F,2000ApJ...539L..13G} signaled a new and fundamental understanding on the nature of black hole and galaxy formation and evolution; 
the $M_\bullet-\sigma_\ast$ relation intimately links the most basic characteristic of a SMBH with an underlying dynamical property of the surrounding galaxy. 
Consequently, the $M_\bullet-\sigma_\ast$ relation, that is typically described by $\log M_\bullet = \alpha + \beta\log(\sigma_\ast)$, has received an extraordinary 
amount of attention. 

The $M_\bullet-\sigma_\ast$ relation is a member of a family of scaling relations all linking $M_\bullet$ to various properties of the host bulge, and possibly the entire gravitational 
mass of the host galaxy. For a comprehensive review of all the scaling relations see \cite{2005SSRv..116..523F}. Despite all these other relations, $M_\bullet-\sigma_\ast$ has 
traditionally been the preferred as, at its time of discovery, it showed a remarkably small scatter; a smaller scatter than for the other scaling relations and a scatter consistent 
with the measurement errors alone. Since then, however, \cite{2003ApJ...589L..21M} and \cite{2007MNRAS.379..711G} have shown that with enough care, scaling relation scatters can be 
comparable especially when using near-infrared luminosities. Regardless of which scaling relation is ``best'', it is important to fully investigate them all as the intrinsic scatters 
themselves contain pivotal information on the precise inter-play between SMBHs and their hosts. 

Reproducing the features of the $M_\bullet-\sigma_\ast$ relation has become an integral part of SMBH and galaxy formation and evolution models. The relation also has several other 
significant applications. For example, the $M_\bullet-\sigma_\ast$ relation may be extrapolated to smaller and larger values of $M_\bullet$, and be used to estimate $M_\bullet$ in systems 
where the black hole sphere of influence radius ($r_h$) cannot be resolved. It is therefore especially important to know the values of $\alpha$ and $\beta$, and their associated scatters, 
to high accuracy. To have doubt in either parameter is to introduce large uncertainties in extrapolated values of $M_\bullet$. This in turn has a fundamental impact on our understanding 
of both black hole and galaxy formation and evolution. 

\section{The $M_\bullet-\sigma_\ast$ Relation}\label{proj}

\begin{figure}
\resizebox{\hsize}{!}{\includegraphics[clip=true]{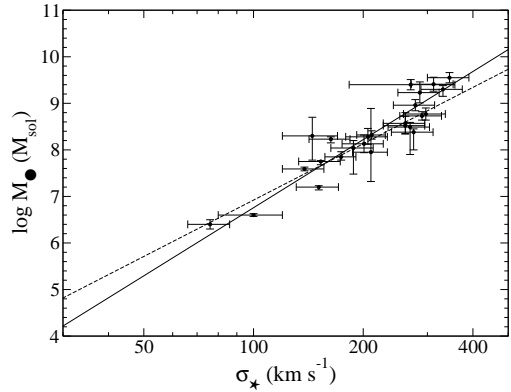}}
\caption{The $M_\bullet-\sigma_\ast$ relation (where $r_h$ is resolved) using the \cite{2005SSRv..116..523F} ($\beta=4.86$, solid line) and 
\cite{2002ApJ...574..740T} ($\beta=4.02$, dashed line) fits.
}
\label{fig:1}
\end{figure}

The $M_\bullet-\sigma_\ast$ relation has had many attempts to fit its slope, zero-point and scatter to the whole galaxy population \citep{2000ApJ...539L...9F,2000ApJ...539L..13G,
2001ApJ...547..140M,2002ApJ...574..740T} in addition to individual families of galaxies that potentially lie off the relation \citep[e.g.][]{2001A&A...377...52W,2005ApJ...633..688M,
2007ApJ...667L..33K}. So far, these individual fits have produced considerably different results ($\beta$ ranges from 3.8 to 4.9, for example). To demonstrate what affect these 
uncertainties have on the importance of the $M_\bullet-\sigma_\ast$ relation, we have used the direct $M_\bullet$ and $\sigma_\ast$ data compiled recently by 
\cite[][Table 2]{2005SSRv..116..523F} to construct Figure~\ref{fig:1}. The fit of \cite{2002ApJ...574..740T} is also plotted as these values are typically used by many authors. Taking 
$\sigma_\ast=20{\rm{km~s^{-1}}}$ (a large globular cluster) the two slopes predict $\log M_\bullet=4.11M_\odot$ or $\log M_\bullet=3.36M_\odot$. Taking $\sigma_\ast=470{\rm{km~s^{-1}}}$ 
(a brightest cluster galaxy) the two slopes predict $\log M_\bullet=9.62M_\odot$ or $\log M_\bullet=10.02M_\odot$.

The aim of the $M_\bullet-\sigma_\ast$ Project is to produce the most reliable estimate of the $M_\bullet-\sigma_\ast$ relation to date. It will couple ground-based integral field 
spectroscopy (IFS) with high spatial and spectral resolution data, from the STIS archive, to estimate $M_\bullet$ in a large sample of galaxies using the latest three-integral, 
axisymmetric, orbit-based stellar dynamical models. The project will also examine the accurate determination of $\sigma_\ast$ within the host bulges. In addition, the project 
will identify and address several key issues that surround the uncertainties in determining $\alpha$ and $\beta$.

\subsection{Estimating $M_\bullet$}

Four methods have been used to {\it directly} estimate the values of $M_\bullet$ defining the $M_\bullet-\sigma_\ast$ relation. Proper motions have been used in the Milky Way 
\citep{1997MNRAS.291..219G,2005ApJ...620..744G} and $\rm H_20$ masers have been used to estimate $M_\bullet$ in NGC~4258 \citep{1995Natur.373..127M}. All other estimates have been 
made using gas and stellar dynamics. Gas dynamics \citep[e.g.,][]{1996ApJ...470..444F,2001ApJ...549..915M}, are relatively easy to model as long as the nuclear disk is completely 
dominated by the gravitational potential of the SMBH. Stellar dynamics  \citep[e.g.,][]{1994MNRAS.270..271V,2002MNRAS.335..517V,2003ApJ...583...92G}, does not suffer from this problem, 
but is a more complex method that requires a high sensitivity instrument. It is difficult to judge the consistency between estimates made by each method as very few galaxies have been 
modeled by more than one method. 

Given a sufficiently sensitive instrument, and a sufficiently elaborate model, then stellar dynamics is the preferred method for determining $M_\bullet$. Indeed, to date 57\% 
of direct $M_\bullet$ estimations have been made using stellar dynamics; it is the dominant method defining the $M_\bullet-\sigma_\ast$ relation. \cite{2004ApJ...602...66V} provide a 
detailed description of an orbit-based model designed to recover the gravitational potential of a system using stellar kinematics. They carry out detailed analytical testing of this model 
as applied to a dataset from M32. \cite{2005ApJ...628..137V} apply the same model to constrain $M_\bullet$ in NGC~205. It is found that a large range in $M_\bullet$ gives equally acceptable 
fits. The degeneracy in the solution is a function of the number of orbits used in the models and the number of constraining data points. 

As M32 and NGC~205 are the only datasets for which $\chi^2$ space in stellar dynamical models has been explored in detail, it is clear that the dominant mass estimates within the current 
$M_\bullet-\sigma_\ast$ relation may include un-investigated issues that could have a fundamental impact on our understanding of the relation itself. The $M_\bullet-\sigma_\ast$ project 
will address these issues in two key ways. Firstly, it will use a wide range of constraining data points by employing IFS. Secondly, it will serially reduce a large family 
of dynamical models, that have a wide range of total orbit numbers, on a high performance computing cluster with a large node number.

It has been repeated by many authors \citep{2002ApJ...578...90F,2003ApJ...589L..21M,2004ApJ...602...66V} that resolving $r_h$ (given by $GM_{\bullet}/{\sigma_\ast^2}$) is a minimum 
requirement for detecting a SMBH. If $r_h$ is un-resolved then estimates of $M_\bullet$ can be tarnished by uncertain contributions from stellar masses. The derivation of $r_h$ was 
first attempted by \cite{1972ApJ...178..371P} based on the ad hoc assumption that in a steady state the stellar density is exponentially linked to the total gravitational potential 
and that $\sigma_\ast$ is constant. The black hole removes stars at small radii (via accretion or ejection, for example) and disrupts this equilibrium. The steady state is restored 
via an increase in $\sigma_\ast$ of stars brought in from larger radii. 

\begin{figure}
\resizebox{\hsize}{!}{\includegraphics[clip=true]{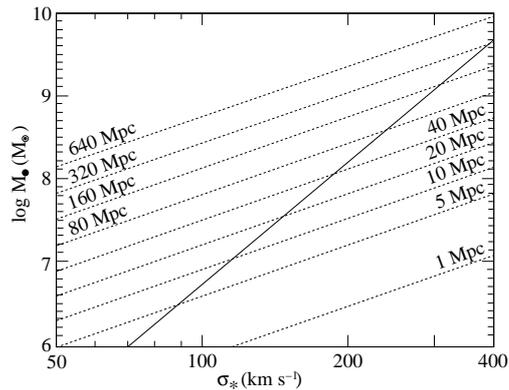}}
\caption{The ability of {\it HST} to resolve $r_h$. Above the dashed lines $r_h$ is resolved and $M_\bullet$ estimates can be made. 
}
\label{fig:2}
\end{figure}

Although the requirement to resolve $r_h$ has been thoroughly argued it must be noted that this creates a somewhat degenerate problem; you are forced to make an assumption about SMBH 
demographics (i.e., that they follow the $M_\bullet-\sigma_\ast$ relation) in order to deduce whether you can detect that SMBH in the first place. Furthermore, you are using the very 
parameters in the $M_\bullet-\sigma_\ast$ relation in the calculation of $r_h$. Nevertheless, with the assumption that SMBHs do follow the $M_\bullet-\sigma_\ast$ relation, we can plot 
$r_h$ within the $M_\bullet-\sigma_\ast$ plane. In addition, we can determine the ability of {\it HST}, from which most $M_\bullet$ estimates are derived, to resolve $r_h$ at a specific 
distance (Fig~\ref{fig:2}).

Two key points are associated with Figure~\ref{fig:2}. Firstly, it does not include the sensitivity, and therefore the SMBH detection efficiency, of the telescope. The relatively small 
aperture of {\it HST} ensures that many STIS orbits are required before a sufficiently large continuum signal-to-noise is reached for stellar dynamical studies. Secondly, it shows that 
even at 16 Mpc {\it HST} is unable to explore a significant area of the $M_\bullet-\sigma_\ast$ plane (below the dashed lines). If there were a population of bona-fide under-massive black 
holes, as suggested by \cite{2007ApJ...663L...5V} for example, {\it HST} would not be able to detect them;  they would not be modeled and the $M_\bullet-\sigma_\ast$ relation would still 
remain to appear universal. In short, {\it HST} has a SMBH ``blind spot''. To unequivocally define the $M_\bullet-\sigma_\ast$ relation the entire $M_\bullet-\sigma_\ast$ plane needs to be 
detectable over a large enough distance to include a large population of host galaxies. 

\subsection{Estimating $\sigma_\ast$}

One of the problems associated with the fitting of the $M_\bullet-\sigma_\ast$ relation concerns the values of $\sigma_\ast$ used. \cite{2000ApJ...539L...9F} transformed $\sigma_\ast$ 
to an aperture of $r_e/8$, where $r_e$ is the host effective radius, while \cite{2000ApJ...539L..13G} use a value weighted by the host luminosity. The use of different values is 
understandable considering there is, as yet, no clearly defined aperture size through which $\sigma_\ast$ should be determined. 

When considering the nature of a LOSVD, the second order of which is usually described as $\sigma_\ast$, one can see that the size and nature of the aperture through which the LOSVD is 
collected has a fundamental bearing on its observed shape. In addition to the intrinsic variation of $\sigma_\ast$ with aperture size \citep{2005ApJS..160...76B}, a velocity field across 
the aperture will rotationally broaden the LOSVD as a function of the aperture diameter. This will produce an over estimate of the intrinsic $\sigma_\ast$. In the case of an 
axisymmetric velocity field, the aperture shape and orientation will also have a bearing; a different value of $\sigma_\ast$ will be derived from, for example, a long-slit aligned 
along the major or minor axis of rotation. IFS can be used to address these concerns.

\section{Summary}

The $M_\bullet-\sigma_\ast$ relation is one of the most valuable correlations in modern galaxy and SMBH formation and evolutionary models. There are, however, concerns in the ways both 
$\sigma_\ast$ and many $M_\bullet$ estimates have been determined. The latest stellar dynamical models have shown that great care must be used to fully explore $\chi^2$ space around 
$M_\bullet$. It is essential that $r_h$ is fully resolved, that there are large number of high quality constraints, and that a large variety of orbit families are employed in the models. 
The most appropriate values of $\sigma_\ast$ may not have been determined as it has yet to be clearly defined. STIS long-slit, and integral field spectroscopy can be used to address all 
these issues and will lead to the most accurate form of the $M_\bullet-\sigma_\ast$ relation to date. 

\begin{acknowledgements}
The author gratefully acknowledges useful discussions with D. J. Axon, A. Marconi, D. Merritt and M. Valluri. 
Support for Proposal number HST-AR-10935.01 was provided by NASA through a grant from the Space Telescope Science Institute, which is operated 
by the Association of Universities for Research in Astronomy, Incorporated, under NASA contract NAS5-26555. 
\end{acknowledgements}

\bibliographystyle{aa}
\bibliography{bbl}

\end{document}